\documentclass[11pt,preprint]{aastex} 

\raggedright
 
 \usepackage{natbib}   

\begin{document}


\shorttitle{Morphology of Open Clusters}
\shortauthors{Chen et al.}

\title{Morphology of Galactic Open Clusters} 

\author{W. P. Chen} 
  \affil{Institute of Astronomy and Department of Physics, National Central 
           University, Chung-Li 32054, Taiwan}
  \email{wchen@astro.ncu.edu.tw}
\author{C. W. Chen}
  \affil{Institute of Astronomy, National Central
           University, Chung-Li 32054, Taiwan}
  \email{awei@outflows.astro.ncu.edu.tw}
\and
\author{C. G. Shu}
  \affil{Shanghai Astronomical Observatory, 
   Chinese Academy of Sciences, Shanghai 200030, and 
   Center for Astrophysics, Shanghai Normal University, Shanghai 200234}
  \email{cgshu@center.shao.ac.cn}

\begin{abstract}
We analyzed the shapes of Galactic open clusters by the star counting technique 
with the 2MASS star catalog database.   Morphological parameters such as 
the ellipticity and size have been derived via stellar density distribution, 
weighed by clustering probability.  We find that most star clusters are elongated, 
even for the youngest star clusters of a few million years old, which are located 
near to the Galactic disk.  The shapes of young star clusters must reflect the 
conditions in the parental molecular clouds and during 
the cluster formation process.  As an open cluster ages, stellar dynamics cause the 
inner part of the cluster to circularize, but the overall radius gets larger and 
the stellar density becomes sparser.  We discuss how internal relaxation process 
competes with Galactic external perturbation during cluster evolution.   
\end{abstract} 

 
\section{Morphology of Star Clusters} 

The way member stars distribute within a star cluster changes as the cluster
evolves.  The {\em initial\/} stellar distribution is governed  
by the structure of the parental molecular cloud, and by how star formation
proceeds.  Sequential star formation, for example, may result in massive 
stars, which are responsible for inducing the formation of 
next-generation stars, to have a birth place markedly different from that of 
low-mass stars.   As the star cluster evolves, the distribution is 
modified by {\em internal\/} gravitational interaction among member stars.  
Subsequently stellar evaporation and {\em external\/} 
disturbances---e.g., Galactic tidal force, differential rotation, and 
encounters with molecular clouds---would alter the spatial structure and  
eventually dissolve the cluster.  
While individual star clusters are good laboratories to study stellar 
dynamics \citep{fri95}, they also serve as test particles to 
probe the local physical (e.g., gravitational field) and chemical 
(e.g., metallicity) conditions.  Open clusters in particular, with their 
wide ranges of age and location distribution, would be valuable tools to 
study the star formation history and chemical evolution of the Galactic 
disk.     

Stars in a globular cluster are known to concentrate progressively
toward the center, more so for massive stars than for low-mass stars.
The density distribution, prescribed by the King model \citep{kin66}, is
understood as a combination of an isothermal sphere (i.e.,
dynamically relaxed) in the inner part of the cluster, and tidal
truncation by the Milky Way in the outer part.
In contrast, open clusters appear irregularly shaped, with member 
stars sparsely distributed.  The youngest clusters must still bear the 
imprint of their formation history, so their structure, when compared 
with that of molecular clouds, may shed crucial light to the 
fragmentation process during cloud collapse.  At the other end, 
the oldest open clusters serve as tracers of the structural and 
evolutionary history of the Galactic disk \citep{fri95}.
Even though some open clusters are known to follow the  
King model \citep{kin62}, and despite some elaborative theoretical 
considerations on the dynamics of open clusters \citep[e.g., ][]{por01}), 
few observational studies have been done systematically until recently 
\citep[e.g., ][]{pan90,nil02}, perhaps due to the complexity 
of the problem arising from the small number of member stars, and the 
lack of comprehensive data of open clusters on large angular scales.  

As early as almost a century ago, \citet{jea16} already considered
the flattening effect of a moving cluster through a general gravitation
field of stars.  \citet{oor79} noticed the uneven distribution of stars 
in Hyades by comparison of the numbers of stars in quadrants about 
the cluster's apparent center.  He concluded that the cluster appeared  
flattened with an axial ratio of about 2 aligned with the Galactic plane.
\citet{ber01} analyzed the surface star density in 3 open
clusters by the star counting method and found elongated shapes parallel to 
the Galactic plane.  These authors furthermore employed wavelet transform 
to bring out possible tidal tails, presumably caused by the Galactic tidal 
field.  

A comprehensive diagnosis of the spatial structure of open clusters
appears elusive because they are loose systems with shallow gravitational
potential, hence lack of organized symmetry in structure, and likewise 
the morphology and shape would be vulnerable to external perturbation.
We would like to learn whether young clusters are mass segregated, and if so, 
to what extent this is due to dynamical relaxation, as opposed
to relic structure in molecular clouds.  To answer 
questions like these, it is desirable to study the spatial structure of the
youngest star clusters, and see how it evolves as a cluster ages and
moves in its Galactic orbit.  

As a pilot study on how a star cluster shapes out of the molecular 
cloud, we \citep{che02} investigated the radial 
star density profiles of 7 open clusters with ages ranging from a few 
million years to a few billion years, based on the 2MASS (Two-Micron
All-Sky Survey) Second Increment Release star catalog.  
The infrared data enabled us to probe the distribution of member 
stars of clusters that are very young and still embedded in molecular clouds.    
Our study indicated that stars, regardless of their masses, tend to 
concentrate progressively toward the center of a cluster, and the 
degree of concentration is higher for luminous (presumably more massive, 
or as binary) stars than for fainter members.  
Such a segregation structure appears to exist in even
the youngest star clusters.  The relaxation time 
$\tau_{\rm relax} \approx (0.1 N/ \ln N)\cdot \tau_{\rm cross}$, where $N$ is the 
number of stars and $\tau_{\rm cross}$ is the crossing time of the system
\citep{bin87}.  
For a typical open cluster $N \sim 10^{3}$, and $\tau_{\rm cross}=R/V$ where 
the size of the cluster $R\sim 2$~pc and the velocity dispersion 
$V\sim 1$~km\,s$^{-1}$, yielding $\tau_{\rm relax} \approx 3 \times 
10^{7}$~yrs.  The youngest open clusters (ages of a few million years) 
therefore have not had time for dynamical relaxation to take place 
efficiently.  The spatial distribution of member stars in the youngest star 
clusters hence is much relevant to the structure in the parental cloud 
out of which the cluster was formed, and to subsequent mass redistribution 
during star (cluster) formation process.  

\citet{nil02} analyzed the radial star density profiles of some 
38 open clusters based on the U.S. Naval Observatory (USNO)-A v2.0 star 
catalog, which in turns was derived from the Digital Sky Survey (DSS) 
plates.  Because no membership information is available, the extent, size, 
shape, or any spatial structure of a particular star cluster is 
estimated in a statistical sense against adjacent Galactic star fields.  
One of the advantages of working with a homogeneous sky-survey database, 
such as the 2MASS or USNO catalog, in addition to the convenient 
availability, is the extended angular coverage to encompass not only 
the entirety of a cluster itself (core, envelope, or possible tails), 
but also sufficient field regions for comparison.  A fair assessment of 
the stellar density fluctuations in comparison fields is crucial in 
the star counting technique.  

Parameterization of the spatial structure of an open cluster by its 
radial distribution alone obviously is not adequate, as 
already pointed out by \citet{nil02}.  Many open clusters 
have highly irregular shapes, often even with no clear centers, 
so circular symmetry cannot be readily assumed.     
Dense molecular cloud cores are shaped on average as a prolate 
Gaussian with an intrinsic axial ratio of 0.54, as a part of 
an evolutionary sequence from filamentary molecular cloud complexes to 
roughly spherically condensed cores \citep{cur02}.  It is therefore 
desirable to represent the stellar distribution by a more sophisticated 
method than a one-dimensional analysis.  This paper summarizes the result 
of our attempt to analyze the morphological shapes of open clusters, 
and how the shaping would evolve in Galactic environments.

We take a probabilistic approach to estimate the stellar surface 
density of stars in the 2MASS point source catalog,  
and represent a star cluster with an ellipsoid.  This allows us to 
investigate not only the structure (concentration, segregation) but also 
the shape (elongation, orientation) of a cluster, and hence by a sample 
of clusters of different ages and environments, to delineate 
the morphological evolution influenced by Galactic dynamics.    
The 2MASS infrared data are free from much of interstellar
extinction, and so would reveal the true shape of a star cluster
more readily than in optical wavelengths.  Some youngest star clusters may 
be seen only in infrared wavelengths \citep{lad03}.  We describe our 2MASS 
sample of open clusters in Section 2, and present the methodology to 
analyze the morphology of open clusters in Section 3.   The results and 
discussions are summarized as the last section.      

\section{Data of Open Clusters} 

\citet{dia02} complied the latest catalog of open clusters, 
which is based on the previous work of \citet{lyn87a} and of 
Mermilliod\footnote{http://obswww.unige.ch/webda}, with some 
updated data on radial velocity, proper motion and metallicity.   
Of the total of more than 1600 entries, about 38\% have distance, age and 
color-excess determinations.  We note that while such a compilation is useful 
for information retrieval, one should exercise caution when deriving 
statistics from the dataset, because the catalog 
is far from completion---some of the entries may not be bona fide stellar 
groups at all, and perhaps a lot more open clusters are yet to be 
discovered.  Severe extinction by
dust near the Galactic plane makes it difficult at wavelengths shortward of
infrared to recover the true shapes, or even their bare existence, of young
star clusters, which as recent studies show may outnumber optically
visible open clusters by a factor of an order \citep{lad03}.

We selected among the first 800 entries in the \citet{dia02} catalog 
open clusters, roughly from RA $0^{\rm h}$ to $12^{\rm h}$ which suffer
less extinction and source crowdedness toward the Galactic center,   
have distance and age determinations available and have angular sizes 
between 3\arcmin and 40\arcmin.  The choice of the angular range, 
somewhat arbitrary, is a convenient 
compromise between spatial resolution and the practical limit of maximal 
data (1 degree field of view) downloadable from the 2MASS web interface.  
We selected those clusters that are rich in density enhancement (by eye) 
and with as complete data coverage in the 2MASS star catalog (All-Sky Data  
Release) as possible, i.e., no nearby bright stars so as to contaminate 
the field.   In addition to bright field stars, young open
clusters often contain hot, luminous members.  A real bright star would 
leave a blank pattern, rendering an incomplete listing in the 2MASS 
database.  Even a moderately bright star 
would cause unreliable astrometric and photometric determinations on  
neighboring stars.   By working with the 2MASS data, our sample 
suffers less brightness contrast between the hottest stars and faint 
members than in the visible wavelengths.  Interpolation of stellar density 
is possible in most cases as long as the contamination is not 
overwhelming (e.g., too bright or too close within the cluster boundary).  
In the study we report here, none of the sample suffers bright-star 
contamination, and a total of 31 open clusters were selected in the 
morphological analysis.  

\begin{figure}
 \plotone{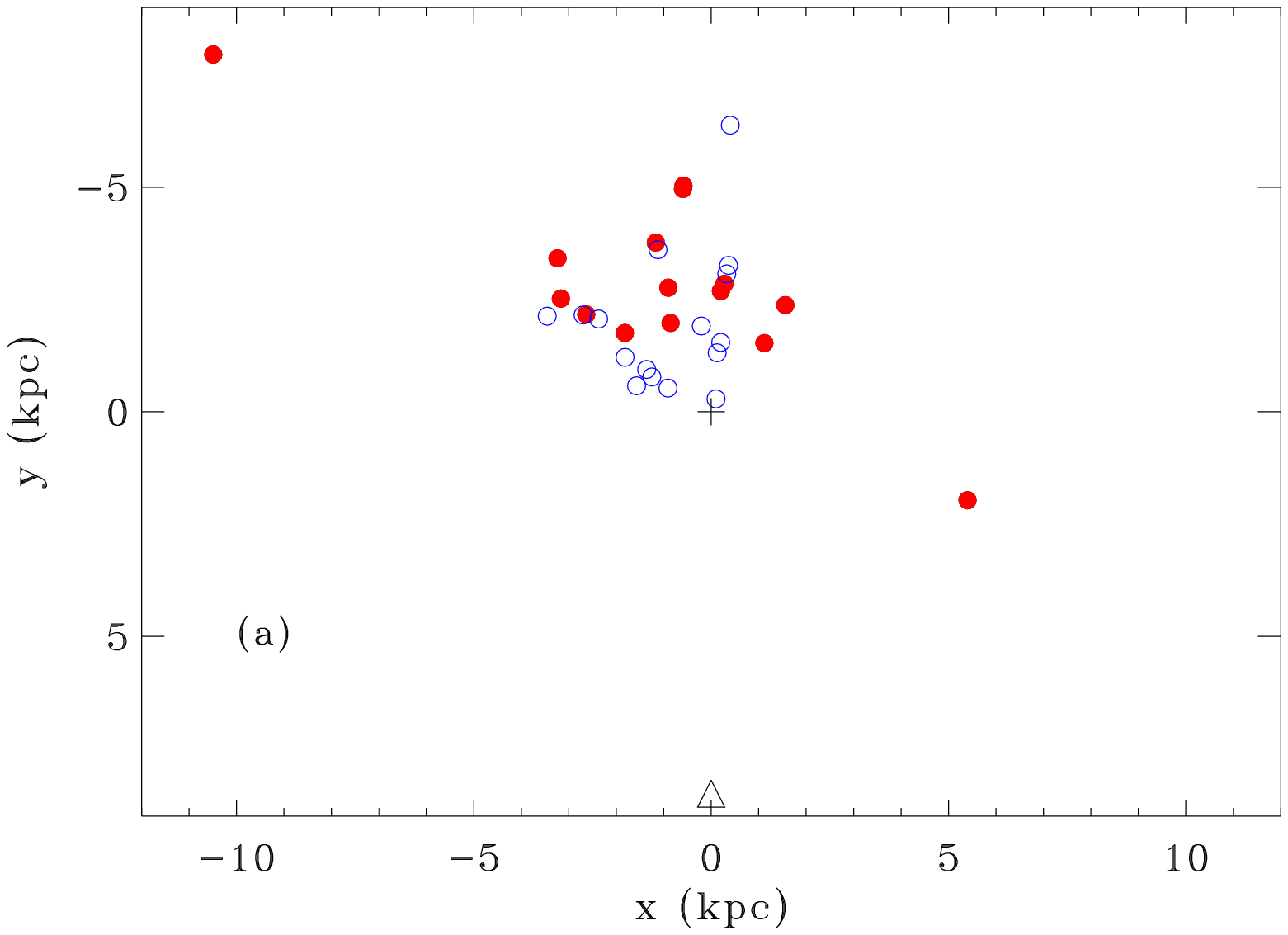} 
 \plotone{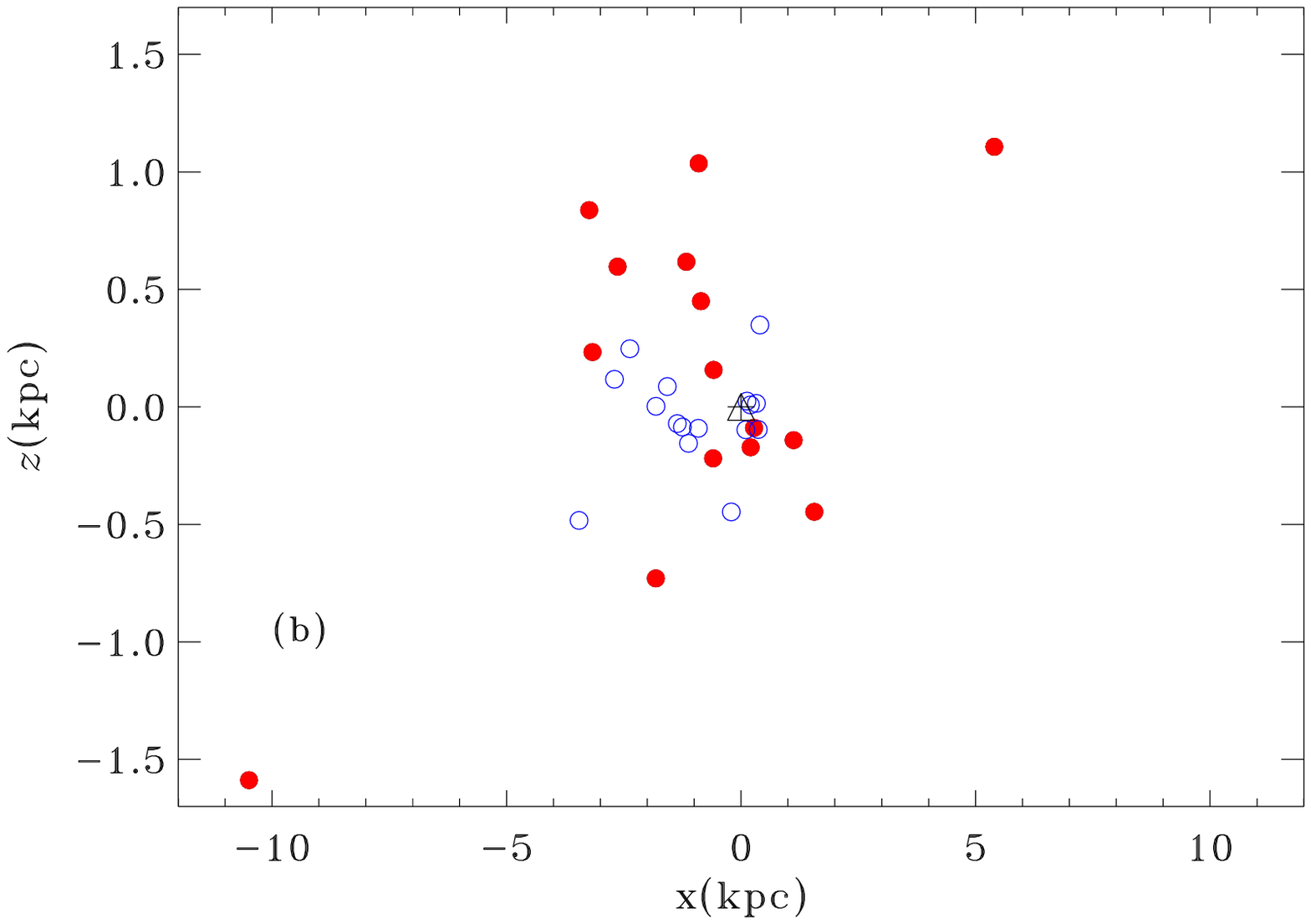} 
\caption{
(a) Face-on view and (b) edge-on view of sample open clusters.  
The sun and Galactic center are marked. 
Young and old clusters (dividing by 800~Myrs, the age of Hyades) 
are denoted, respectively, as open and filled circles.  
To the upper left near $(x,y) \approx (-10.5~{\rm kpc}, -8~{\rm kpc})$ 
is Tombaugh~2, which is among the most distant open clusters from 
the Galactic center and from the Galactic plane \citep{adl82}.  
Located in the inner solar-circle near 
$(x,y)\approx (5.4~{\rm kpc},2~{\rm kpc}) $ 
is NGC\,6791, one of the oldest and most metal-rich 
open clusters \citep{cha99,fri02}. 
The edge-on view shows clearly that old open clusters occupy a wider range 
above and below the disk plane than young systems.
         }
      \label{fig:xyxz}
\end{figure}

Our sample has no obvious additional selection effects, other than 
to avoid the Galactic center where the open cluster catalog \citep{dia02} 
itself may be highly incomplete.  Figure~\ref{fig:xyxz} shows the pole-on 
and edge-on Galactic distributions of our star cluster sample.   
Clusters younger and older than 800~Myrs are marked differently.   
It is noted that the majority (29) 
of our sample are in the direction of Galactic anti-center, a consequence of 
our sample selection from half of the \citet{dia02} catalog.  It is 
seen that old open clusters have a larger average scale height above 
the Galactic plane than young star clusters do.  

\section{Methodology } 

\subsection{Statistical Membership}

In the absence of membership information on individual stars,
we estimate the structure of a star cluster by a probabilistic
star counting technique, i.e., the boundary, shape, size are
all determined in a statistical sense.  In essence, the
degree of clustering of neighbors around any star gives
a measure of the likelihood of that star being in a cluster 
environment.  One defines for each individual star $i$ the clustering
parameter,  

\begin{equation} 
  P_i = (N_t - N_f) / N_t = 1 - N_f/N_t, 
  \label{eq:prob}
\end{equation}

for which $N_t$ is the total number of neighboring stars
within a specified angular size (a ``neighborhood aperture''  
centering on the $i$th star) and $N_f$ is the 
average number of field stars within the same aperture. 
The number of field stars can be estimated by 
$ N_f = \Sigma_f \times \pi r_p^2$
where $\Sigma_f$ is the surface number density of field stars, 
estimated in regions away from the apparent star cluster 
and $r_p$ is the radius of the neighborhood aperture.  

We see that the clustering parameter $P_i$ gives a measure of 
local enhancement of stellar density, whose value   
ranges from $P_i \sim 0$ in a field region (for which 
$N_t \sim N_f$) to 
$P_i \sim 1$ near a rich cluster ($N_t \gg N_f$).  In 
other words, $P_i$ behaves very much like a probability 
of cluster membership.  Our probabilistic method to analyze 
membership in a star cluster is similar to that used by \citet{dan90}.

Note that the choice of $r_p$ should not be arbitrary.  
If $r_p$ is too small, the uncertainty in $P_i$ will be large 
due to small-number statistics.  On the other hand, if $r_p$ is 
too large, the intrinsic structure of the
cluster will be smoothed out and detailed structure information 
is no longer available.  

We describe below our procedure to select an appropriate aperture size 
$r_p$ from the surface density of field stars $\Sigma_f$.  Once  
both $r_p$ and $\Sigma_f$ for a cluster are determined, we
can estimate the membership probability for each star by 
Eq.~\ref{eq:prob}. 

\subsection{Surface Density of Field Stars}

The cluster membership probability defined above is based on a measure
of enhancement of local stellar density compared to field stars.  A fair 
estimate of the density of field stars therefore is prerequisite.   
For each cluster, we use regions away ($\gtrsim 20\arcmin$) 
from the central cluster to estimate the mean number density 
of field stars.  

\begin{figure}
 \plotone{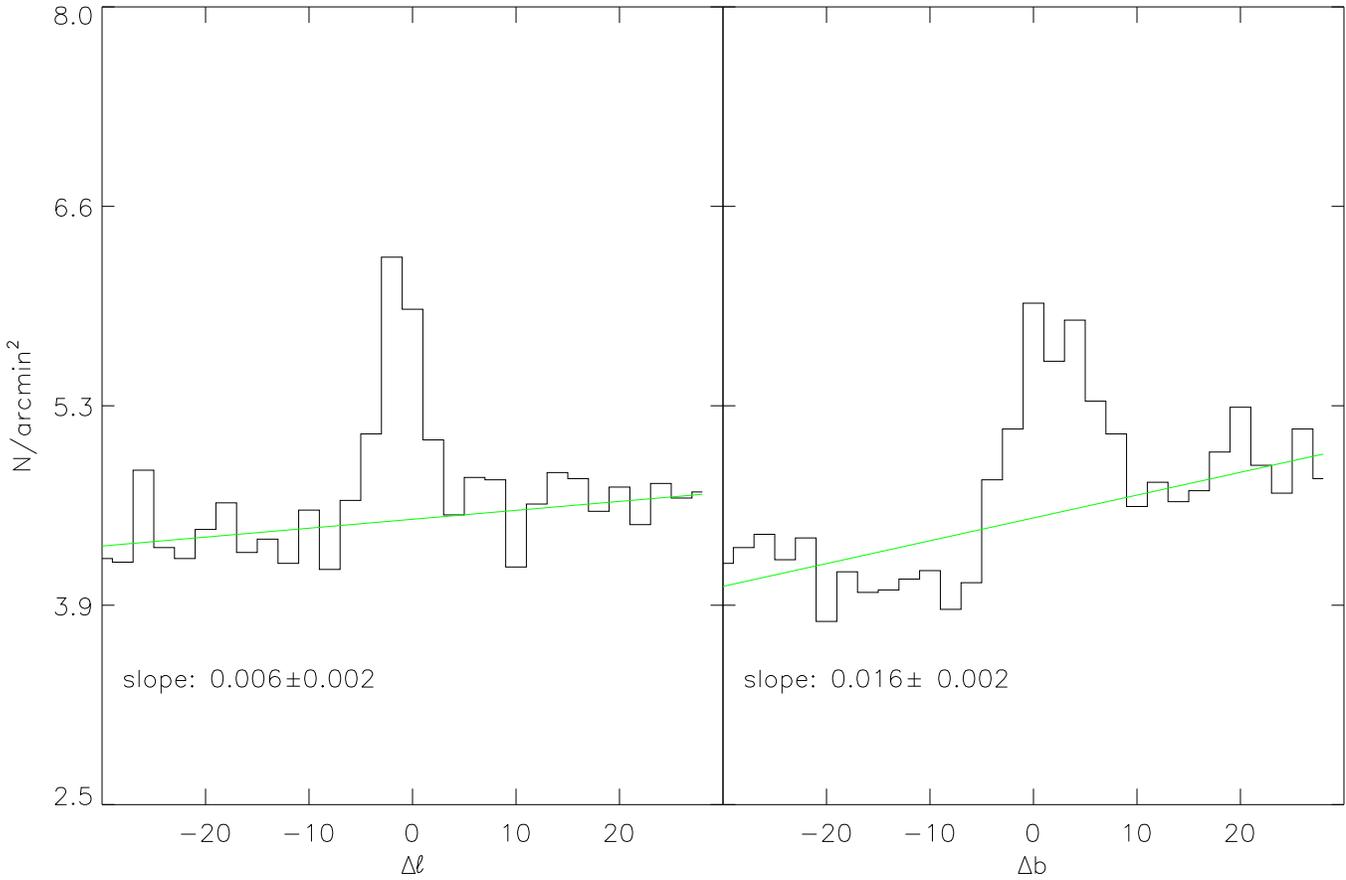} 
 \caption{The number density of field stars exhibits gradients, often  
   pointing toward the Galactic plane.  In the example shown here 
   for NGC\,1893, the projected gradient across a 1~deg squared 
   field, which amounts to $\sim$~2~sq.\,arcmin$^{-2} $ per degree, 
   is significantly smaller in the Galactic longitude than 
   along the latitude direction. } 
 \label{fig:bkgr}
\end{figure}

Since our analysis of the cluster parameters is based on the statistics 
(star density, aperture size) in the surrounding field regions, obviously 
a homogeneous distribution of field stars must be assumed in order for 
our technique to work.  We note actually in almost every case we encountered,
the distribution of field stars is {\em not} homogeneous, due to the
general stellar gradient vertical to the Galactic disk.
Figure~\ref{fig:bkgr} shows such an example for which
a density gradient is seen in both the RA and the DEC projections, with  
the overall gradient vector pointing toward the Galactic disk, as expected.
We also find that, not surprisingly, the density gradient is higher 
for a line of sight with a lower Galactic latitude.     
In our analysis we removed a flat surface density as background for individual 
star clusters, but empirically found this kind of density gradient---though 
potentially useful for study of Galactic disk stellar populations---to have  
little effect on our morphological results because the density differs no 
more than a few percents across the fields of our star clusters.  

The surface number density of field stars $\Sigma_f$ is computed by 
counting the number of stars in a certain sky area.  Due to the discrete nature 
of individual stars, even for a uniform star field, $\Sigma_f$ would 
approach a constant only when the sampling sky area is large enough to include 
a sufficient number of stars.  Otherwise, when the sampling size is smaller 
than about the average angular separation of stars, large fluctuations result.  
As the minimum, one would demand a sky area to have a signal-to-noise 
of $\sqrt{N_f} \gtrsim 3$ against Poisson fluctuations, or $N_f \gtrsim 10$ 
field stars to determine accordingly the optimal neighborhood aperture size
$r_p$.  In our analysis we take $N_f=50$ (i.e., $S/N \sim 7$) and 
select the corresponding $r_p$ for each cluster.  
In general $r_p$ is on order of a few arcminutes.       

\subsection{Cluster Shape and Morphology}

With each star now being represented by a membership probability, the surface
density of cluster member stars is then the sum of the clustering parameter of 
every star within each sky-coordinate grid (with area $\Delta S$); that is, 
the effective number density of member stars is,   

\begin{equation} 
 \Sigma_{s}=\sum P_{i} / \Delta S
 \label{eq:memden}
\end{equation}

Obviously in field regions, $\Sigma_{s} \rightarrow 0$.  
The morphology of a cluster is prescribed by density contours, 
at both the one-third level of the maximal density ($\Sigma_{\rm max}/3$), 
and at the boundary, defined as where the density drops to 2--3 times the 
background fluctuations.  In the extreme case, e.g., in NGC\,2567, for which 
the star cluster is very sparse, a mere 1-$\sigma$ outer boundary 
was used.   Effectively the inner and outer ellipses trace, 
respectively, the core and the halo (or the corona as termed by some
researchers) of a star cluster.  The density contours are least-squares 
fitted with ellipses to obtain the eccentricities of the inner and 
outer ellipses, and the corresponding sizes (average of the semimajor 
and semiminor axes).  We use the flattening parameter to quantify 
the shape of an ellipse.  The flattening
parameter (or oblateness\footnote{http://mathworld.wolfram.com/})
is defined as $f=1-(b/a)$, where $a$ and $b$ are respectively the semimajor and 
semiminor axes of an ellipse, so $f$ is related to the eccentricity $e$ by
$ e^{2} = 1 - (b/a)^{2} = 1 - (1-f)^{2}$.  For $b/a$ very close to 0 (highly flattened) or 
1 (our case), the flattening parameter $f$ is more discriminative than the 
eccentricity itself.     

The uncertainties in the determination of the flattening are estimated by 
Monte Carlo simulations of star clusters of different 
shapes.  For typical parameters of the open clusters in our sample, i.e., 
with $\sim 3$ times enrichment of stellar number density of cluster 
members with respect to the field, the uncertainties in the flattening for 
the outer boundary and for the core (1/3 maximum) are 
$\delta f_{\rm out} \sim 0.18$ and 
$\delta f_{1/3} \sim 0.12$ for a spherical cluster (i.e., $b/a=1$), and 
are  $\delta f_{\rm out} \sim 0.12$ and
$\delta f_{1/3} \sim 0.06$ for an elongated cluster with $b/a=0.5$. 
Obviously, the richer the cluster, the smaller the uncertainties.  
For instance, for a spherical globular cluster with 100 times 
density enhancement, $\delta f_{\rm out} \sim 0.06$, and 
$\delta f_{1/3} \sim 0.02$.   

In addition to the shape (flattening) and size, other parameters, 
such as the position angle of the ellipse, can also be obtained.  
We illustrate in Fig.~\ref{fig:shapes} examples of two clusters with  
different morphology, one of relatively round shape (NGC\,2414) 
and the other of elongated shape (NGC\,1893).  These two 
clusters will be compared in details in the next section.   

We list the results of individual clusters in our sample in 
Table~\ref{tab:results}, where the first column gives the name of 
the cluster, the second column is the Galactic coordinates in degrees, 
and the next three columns are the heliocentric distance, 
height from the Galactic disk, and the age of each star cluster, 
with the distance and age taken from \citet{dia02}.  
Columns 6 to 11 list, respectively, the flattening parameters and 
corresponding sizes (average of semimajor and semiminor axes) 
derived from the density contours of member stars.  The last column
gives the total number of member stars, by summing the total membership
probability of a cluster.    

\section{Morphological Evolution of Open Clusters}

Two kinds of dynamical effects act on, and influence the morphology
of, an open cluster.  The first is internal interaction of two-body
relaxation due to encounters among member stars.  This leads to 
stellar distribution in spherical shape, ever denser toward the
cluster center.  At the same time, low-mass members may gain enough
kinetic energy through the encounters and get thrown out of the system
(i.e., stellar ``evaporation'').  The other dynamical process is external 
interaction, including tidal force due to Galactic disk or giant 
molecular clouds, and differential rotation especially for a cluster 
located in the inner disk region.  These disruptive effects 
tend to make a star cluster elongated in shape, with the outer parts 
particularly vulnerable.

\begin{figure}
 \plotone{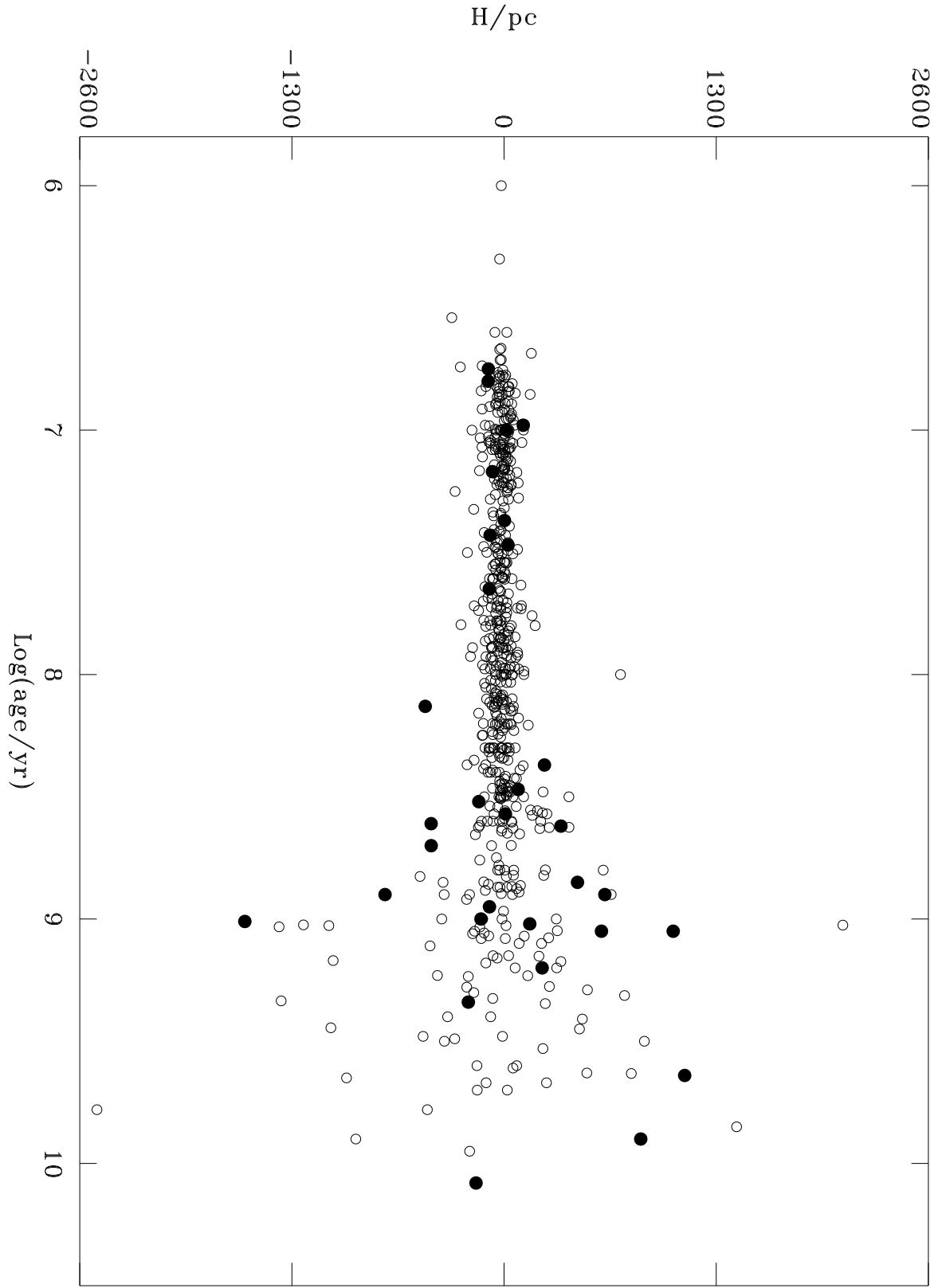} 
 \caption{ The age versus height from Galactic disk for open clusters in 
the \citet{dia02} catalog.  Filled circles are the 31 open clusters
studied in this paper.   
        \label{fig:zt}
         }
\end{figure}

Tidal disturbance on a star cluster is stronger when closer to the Galactic 
disk plane.  Fig.~\ref{fig:zt} shows the vertical heights 
from the Galactic disk versus the ages of the open clusters in the 
\citet{dia02} catalog, of which our sample, separately marked, is 
a subset.  As can be seen, young open clusters tend to reside close  
to the disk where molecular clouds, from which the star clusters were 
formed, are distributed \citep*{tad02, che03}.  
If open clusters are separated into two age
groups, old and young, with the dividing age, somewhat arbitrary, of that of
the Hyades, 0.8~Gyr \citep{phe94,che03}, the scale heights are 354~pc (old)
and 57~pc (young), respectively \citep{che03}, based on the data
compiled by \citet{dia02}).   \citet{tad03} included additional consideration
of the galacto-distances of the clusters and the results support the
previous assertion \citep{lyn87b} that old clusters seem to distribute
at larger scale-heights in the outer parts of the Galaxy than in the 
inner parts.  Obviously, only star clusters away from the inner disk 
regions---where the tidal force from giant molecular clouds plays a 
major disruptive role in the structure or even the existence of a 
star cluster---would have survived on Galactic dynamical time scales 
\citep{jan94}.    

\begin{figure}
 \plotone{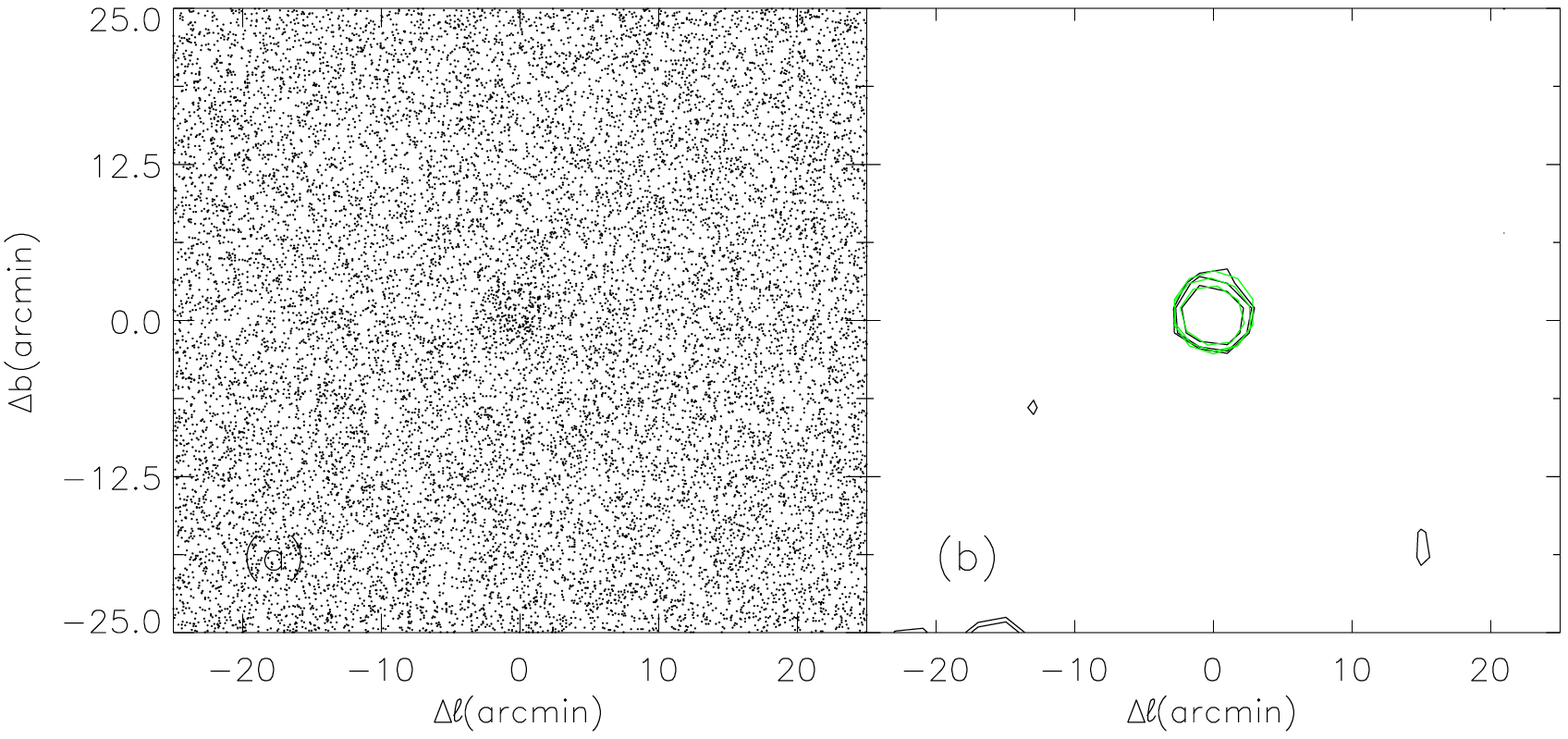} 
 \plotone{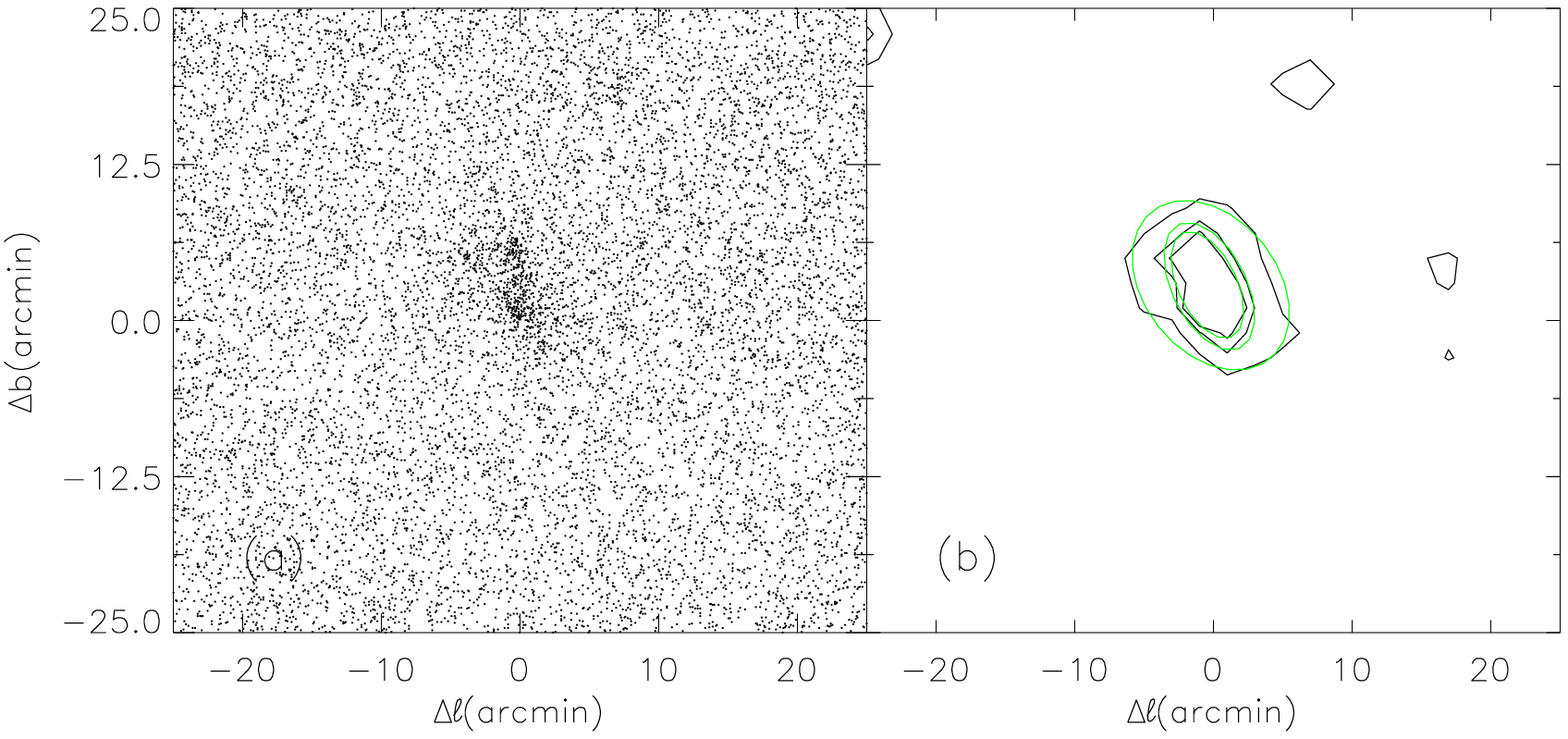} 
\caption{Example of (top) a spherically shaped open cluster, 
NGC\,2414 and (bottom) an elongated open cluster, NGC\,1893.  
In each case, all stars brighter than 15.6~mag, the 
$\sim 3-\sigma$ limit of 2MASS Ks band, are plotted on 
the left panel and the right panel shows the isodensity contours 
of cluster member stars, superimposed with the ellipses fitted to the   
core (1/3 maximum number density) and to the outer boundary 
of the cluster (see text.)   
         }
 \label{fig:shapes}
\end{figure}

Comparison between the two open clusters in Fig.~\ref{fig:shapes} is
informative.  NGC\,2414 is relatively poorly studied, perhaps because
of its paucity and small angular extent.  Much of the literature
about this open cluster can be traced back to the photometric
measurements in \citet{vog72}, based on 10 member stars.  In comparison,
NGC\,1893 is prominently stretched toward the disk plane.   
The cluster is associated with bright nebulosity
and dark clouds, but does not seem to show positional variation of color
excess $E(\bv)$ across the cluster \citep{yad01}.  The extinction effect
would certainly be even smaller in the 2MASS 2~\micron\, data.
Its elongated shape (or two subgroups) thus should be 
inherent to the stellar distribution 
within the cluter, rather than caused by extinction variation.

These two clusters share some similarities, namely both being relatively young,
$\log {\rm (age/yr)}\sim 7$, and close to the disk plane, with NGC\,2414
at $\ell \sim 231\arcdeg$, $b \sim +2\arcdeg$, and NGC\,1893 
at $\ell \sim 174\arcdeg$, $b \sim -2\arcdeg$.  They however contrast 
greatly in shape, size, and apparent richness.  NGC\,2414 is round (flattering 
$\sim 0.1$ throughout the entire cluster), small, and contains some 74 member 
stars within its derived radius $\sim 3\arcmin$.  NGC\,1893 on the 
other hand is oval, twice as large in angular extent, 
and encompasses 645 member stars.  NGC\,1893 therefore has a much 
stronger gravitational binding than NGC\,2414 against external 
disruption.   What we see now in NGC\,2414 may well be its remnant 
cluster core.    

\begin{figure}
 \plotone{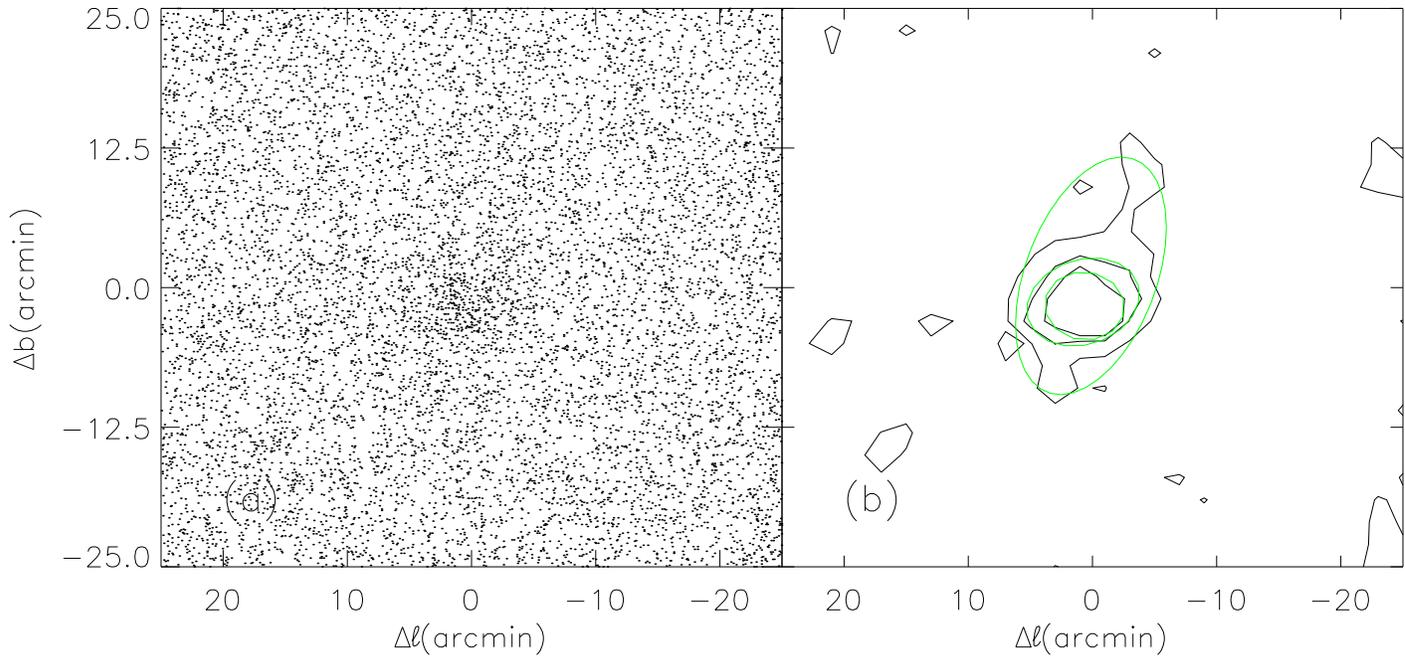} 
 \caption{Morphology of the old open cluster Berkeley~17 
     ($\ell\sim 176\arcdeg, b\sim -4\arcdeg$) shows  
     an upward tail toward Galactic disk.  The outer boundary 
     is duly fitted with an ellipse, obviously not appropriate here. 
           }
 \label{fig:be17}
\end{figure}

Such a stripping off of cluster halos is not uncommon.  For example, 
Berkeley~17 (=C0517+305), the oldest open cluster known 
\citep{sal04}, has a protrusion manifestly pointing 
toward the Galactic disk (Fig.~\ref{fig:be17}), with a projected 
extent comparable to the cluster's radius, $\sim 6$--7\arcmin, or 
about 5~pc in projected length assuming a 2.7~kpc distance \citep{dia02}.  
The enhancement ("corona") has already been inferred from radial 
stellar density distribution, and from the color-magnitude diagram 
of stars away from the nominal cluster region \citep{kul94}. 
Our analysis brings up clearly the tail and its geometry, and further 
hints on an associated antitail, which typifies tidal distortion. 

\begin{figure}
 \plotone{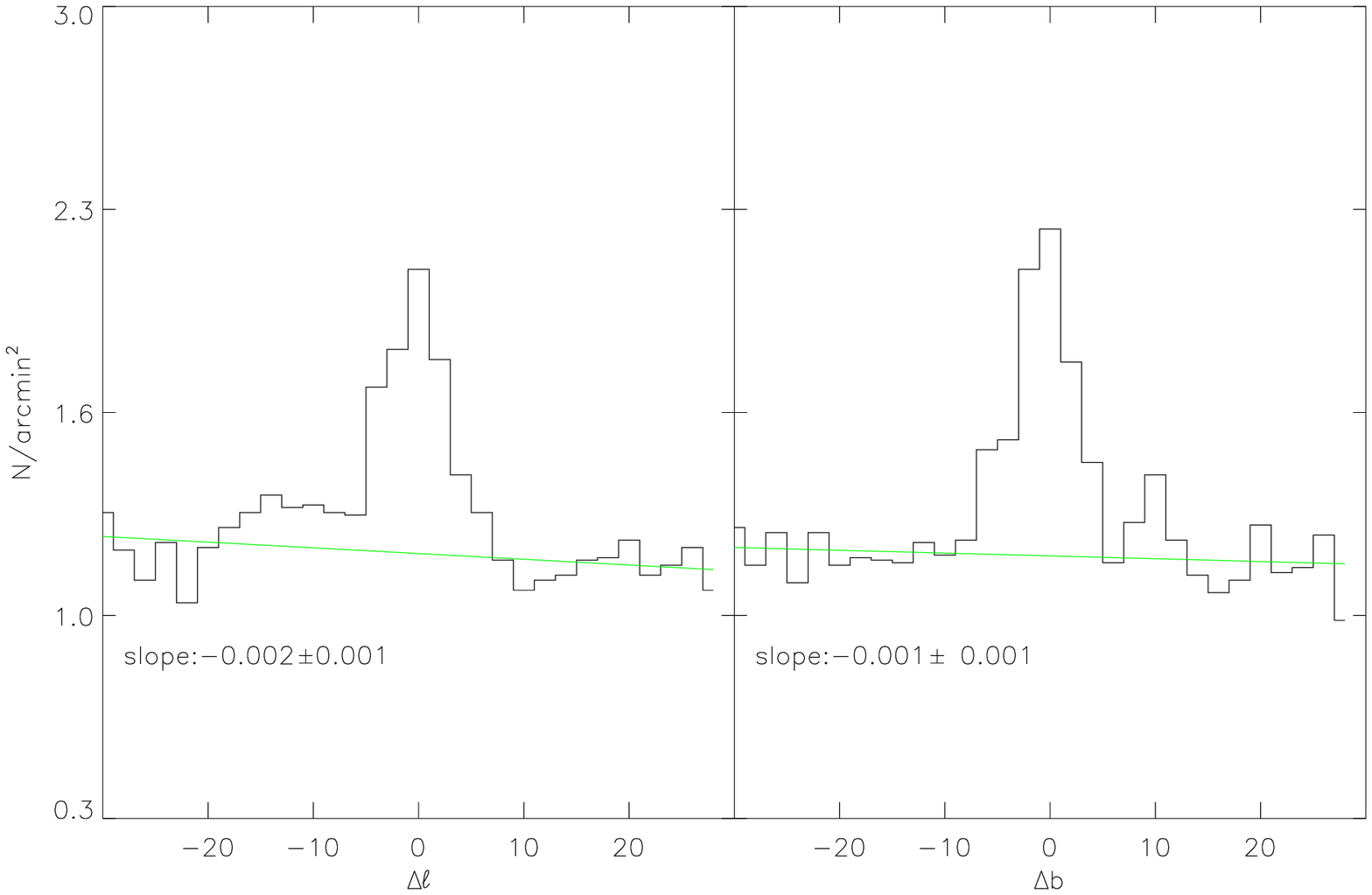} 
 \plotone{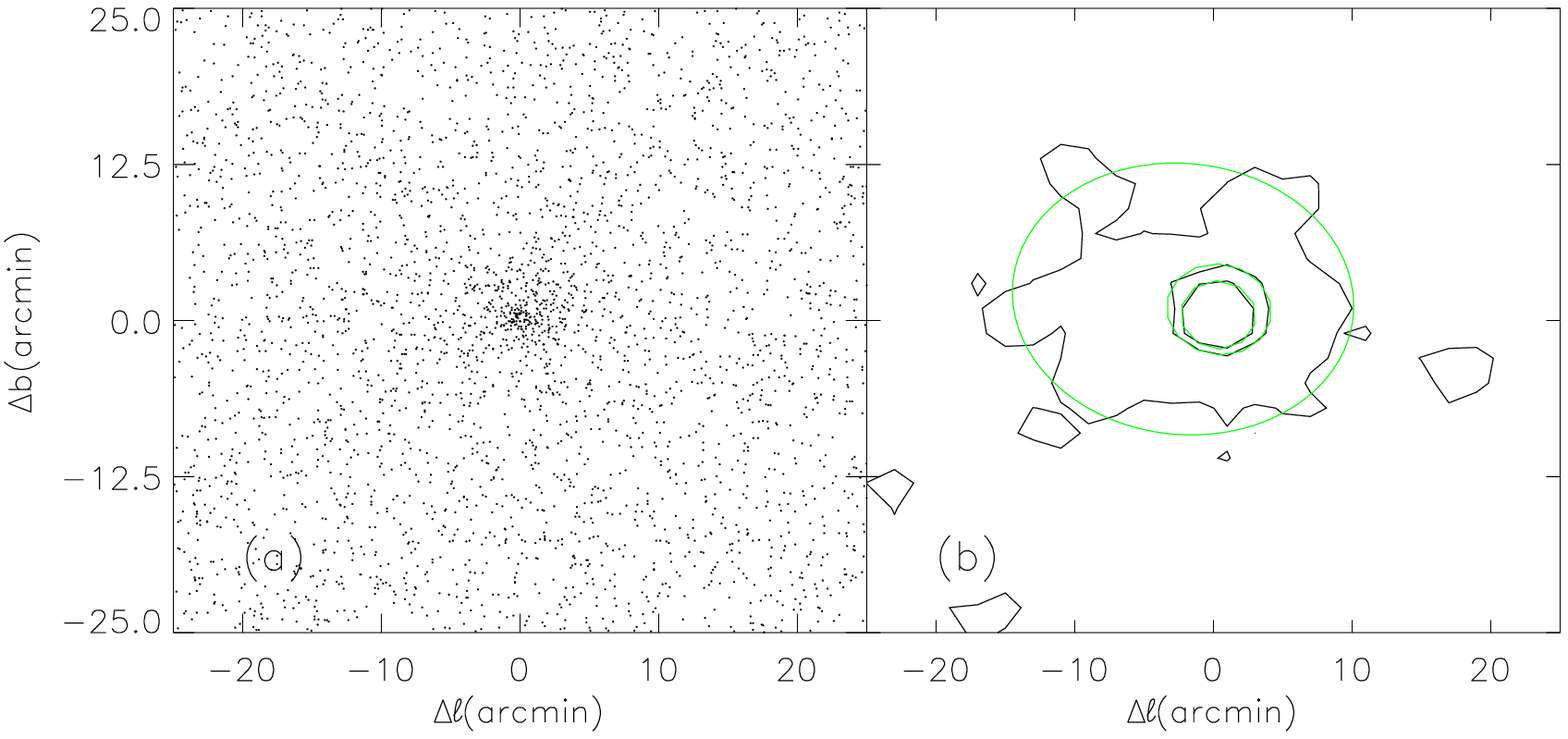} 
 \caption{NGC\,2420, being high above the Galactic plane, 
has a relatively uniform background stellar density in both the 
longitudinal and latitudinal directions, as seen in the top panel.  
The lower panel has the same format as that in Fig.~\ref{fig:shapes} 
and shows  a rather spherical core and a loosely defined outer 
boundary for this old cluster.   
          }
 \label{fig:ngc2420}
\end{figure}

The old, metal-poor, and large-height cluster NGC\,2420 \citep{fri02},
serves as a good showcase for interplay between internal stellar dynamics
and external disturbances.  This cluster is nominally listed in
\citet{dia02} to have an angular diameter of 5\arcmin, at a distance of
3085~pc, and a logarithmic age of 9.048 years, apparently taken from 
WEBDA.  We adopt this distance and those of other star clusters in 
our studies from the \citet{dia02} catalog as a homogeneous source of 
input data, but note that this cluster may be considerably closer, 
at 2.28~kpc \citep{jan94}.  Errors in distance estimation---which often 
turn out to be quite significant among open clusters---would not 
affect the shape determination, but obviously
would influence the results of any statistical analysis.
Using star counting on a Palomar Sky Survey plate, \citet{leo88}
already noticed a much larger extent, to at least 20\arcmin, from the
apparent center of NGC\,2420.  Our analysis of the 2MASS data
shows a negligible gradient in the field star density toward NGC\,2420
(Fig.~\ref{fig:ngc2420}).  The cluster itself is determined to have
an inner (1/3 maximum) and outer (3-$\sigma$ sky) ellipses with
flattening of $f_{1/3} \sim 0.06$, and $f_{\rm out} \sim 0.12$,
respectively.  Average sizes of  
$r_{1/3} \sim 3\farcm7$ and $r_{\rm out} \sim 11\farcm6$
were obtained.  For NGC\,2420, an aperture size $r_p=3$\farcm6 
has been used to estimate the
field star density, $\Sigma_f \sim 1.2\pm 0.1$~arcmin$^{-2}$, from which
the surface density of member stars is then derived according
to Eq.~\ref{eq:memden}.  At the 2MASS limit of Ks=15.6~mag, by 
summing up all $P_i$s (Eq.~\ref{eq:prob}), 
there are a total of 468$\pm$22 member stars.  This is to be compared with
685$\pm$27 within a radius of 20\arcmin\ up to the completeness of 
photographic $\sim$19.5~mag studied by \citet{leo88}, which  
covers a larger sky area with deeper stellar photometry than 
the 2MASS data we have used.  We may be witnessing the disintegrating 
process of the outer part of NGC\,2420.     

Globular clusters are known to
have elliptical shapes \citep{whi87}, likely due mainly to
their overall rotation, rather than to Galactic tidal interactions
\citep{kin61,mey86}.  \citet{kon90} compared globular clusters in the Large
and Small Magellanic Clouds, and found that in virtually all cases,
the inner parts are more elliptical than the outer parts.  This is
understood as the diminishing effect of rotation from the inner to the outer
parts of a globular cluster.   These authors also found that the SMC
globular clusters are more elliptical than those of the LMC, which
in turn are more elliptical than those in our galaxy, and
that the outer shapes are somewhat flatter for younger systems
\citep{kon91}.

Open clusters are also recognized to have elongated shapes
but the orientation of the flattening, however, cannot be accounted for
by Galactic tides alone \citep{jef76}.  This may be because a giant
molecular cloud, local to a particular open cluster, plays a more
influential role in shaping the cluster than a general Galactic disk
potential \citep{dan96}.  Our analysis shows that
even the youngest open clusters (several Myrs old) have very elongated
shapes, with $f \sim 0.5$.   This is to be compared with the
inferred mean intrinsic axial ratio of 0.54 for the observed shapes of
molecular cloud cores \citep{cur02}.  The youngest star clusters hence 
appear by and large to have inherited the morphological shapes from the
prenatal molecular clouds.  Because of the low volume stellar number
density of an open cluster, internal dynamics never becomes dominant
so as to sphericalize the system \citep{por01}, in contrast to
the case in globular clusters.

\begin{figure}
 \plotone{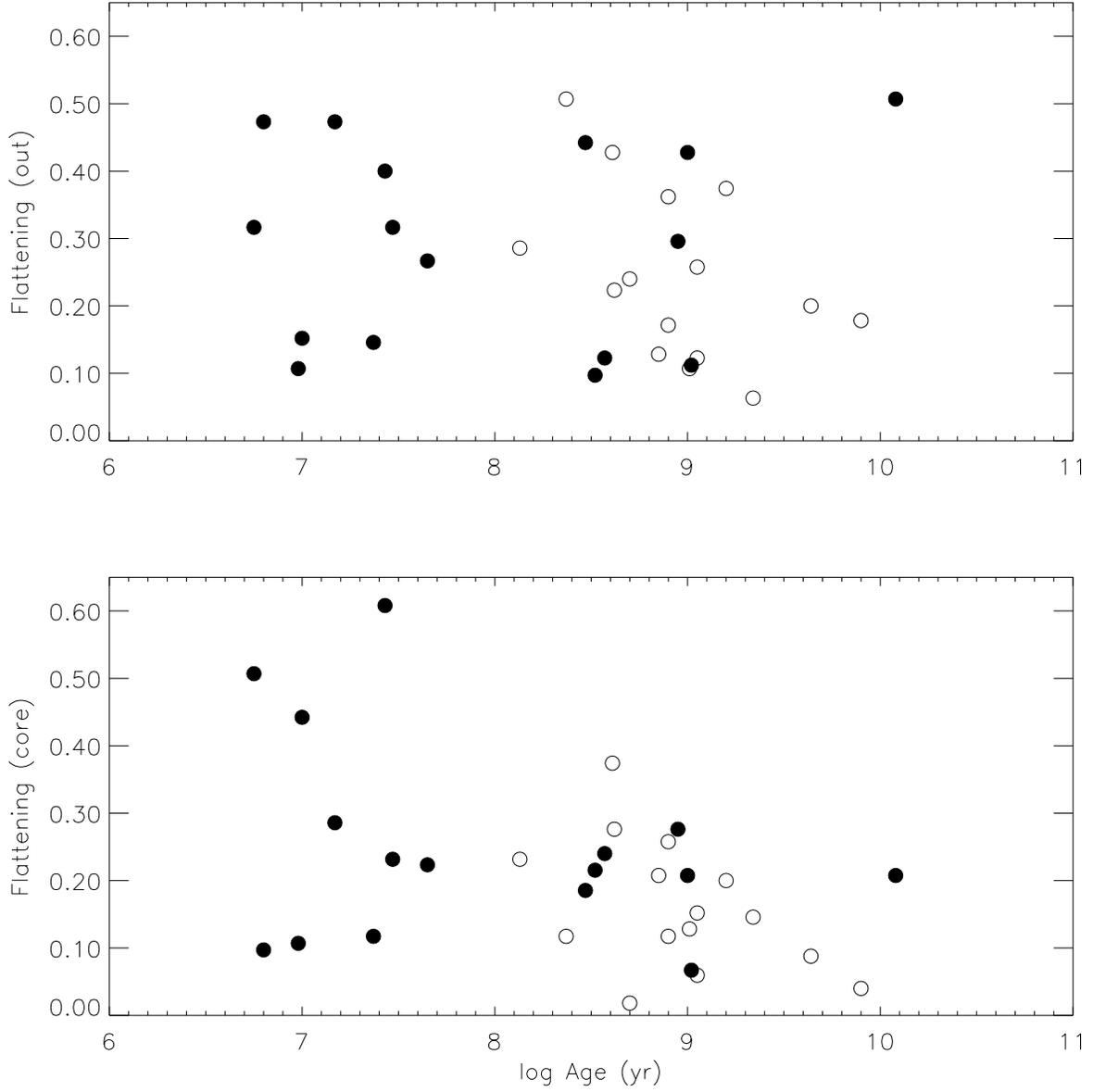} 
 \caption{ The flattening versus age for (top) the outer 
           boundary and (bottom) the inner part of a cluster.  Filled circles  
           denote half of the sample of the clusters which are 
           relatively close (less than $\sim 170$~pc) to the Galactic plane.  
            }
 \label{fig:ft}
\end{figure}

\begin{figure}
 \plotone{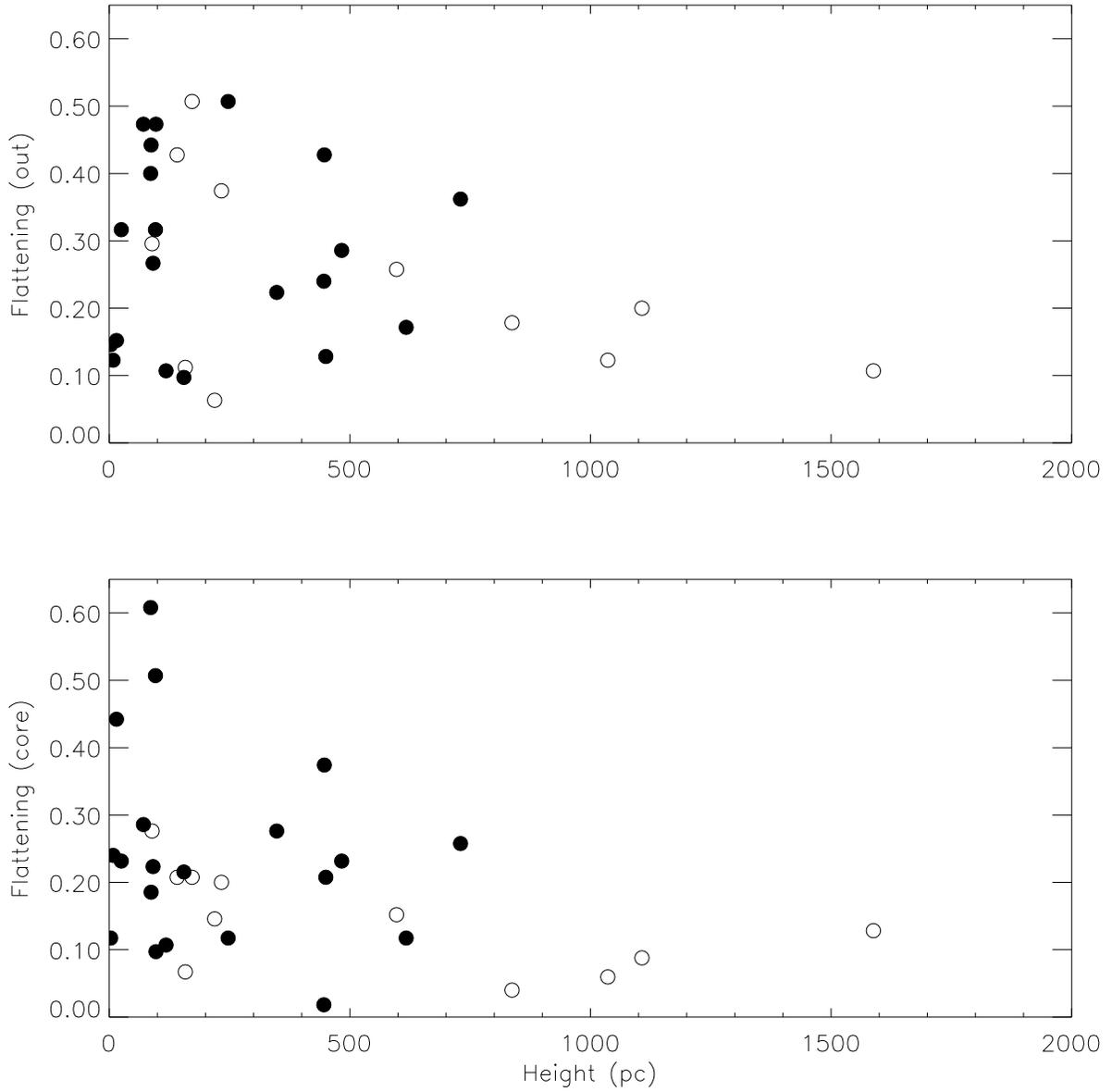} 
 \caption{ The flattening versus the height from the Galactic disk for 
           (top) the outer boundary and (bottom) the inner part of a cluster.  
           Filled circles are for younger star clusters in the sample.       
           }   
 \label{fig:fz}
\end{figure}

Fig.~\ref{fig:ft} and Fig.~\ref{fig:fz} summarize our analysis of 
the shapes of the open cluster sample.  In Fig.~\ref{fig:ft} the shapes of 
the cluster core (diagnosed with $f_{1/3}$) and of the outer 
boundary ($f_{\rm out}$) are shown as a function of age.  One sees that 
most open clusters are elongated in shape.  Because what we see is 
the {\it projected\/} shape, the clusters can only actually be flatter.
The number of our sample of open clusters 
is small (31), yet it is tantalizing to note that while the shape of 
the outer boundary remains unconstrained with age, the core tends to 
circularize as a star cluster ages.  If the sample is further divided 
to two groups according to the height from the disk plane $|H|$ by 
the median value of the sample $\sim 170$~pc, one sees clearly that 
the large-height group (denoted by open circles in Fig.~\ref{fig:ft}) 
evinces a noticeable tendency of rapid spheridalization both in the 
core and in the halo.  Low-lying clusters remain elongated from birth 
to date, prominently so for the outer halos, which are the most 
vulnerable to disruption.  
Fig.~\ref{fig:fz} shows how the flattening of star
clusters varies with the height from the disk plane, for young
(filled circles) and old (open circles) systems.  The youngest star
clusters close to the disk clearly manifest the most elongated shapes.
The tendency toward spherical shaping away from the disk, in particular
in the cores of old systems, demonstrates again the working of internal
dynamical relaxation.
The flattening shapes of young clusters 
are primarily inherited from cluster formation process, but after some  
$\sim 10^{8}$~yr or so, internal stellar dynamics become effective 
in shaping the core of a cluster.  

\begin{figure}
 \plotone{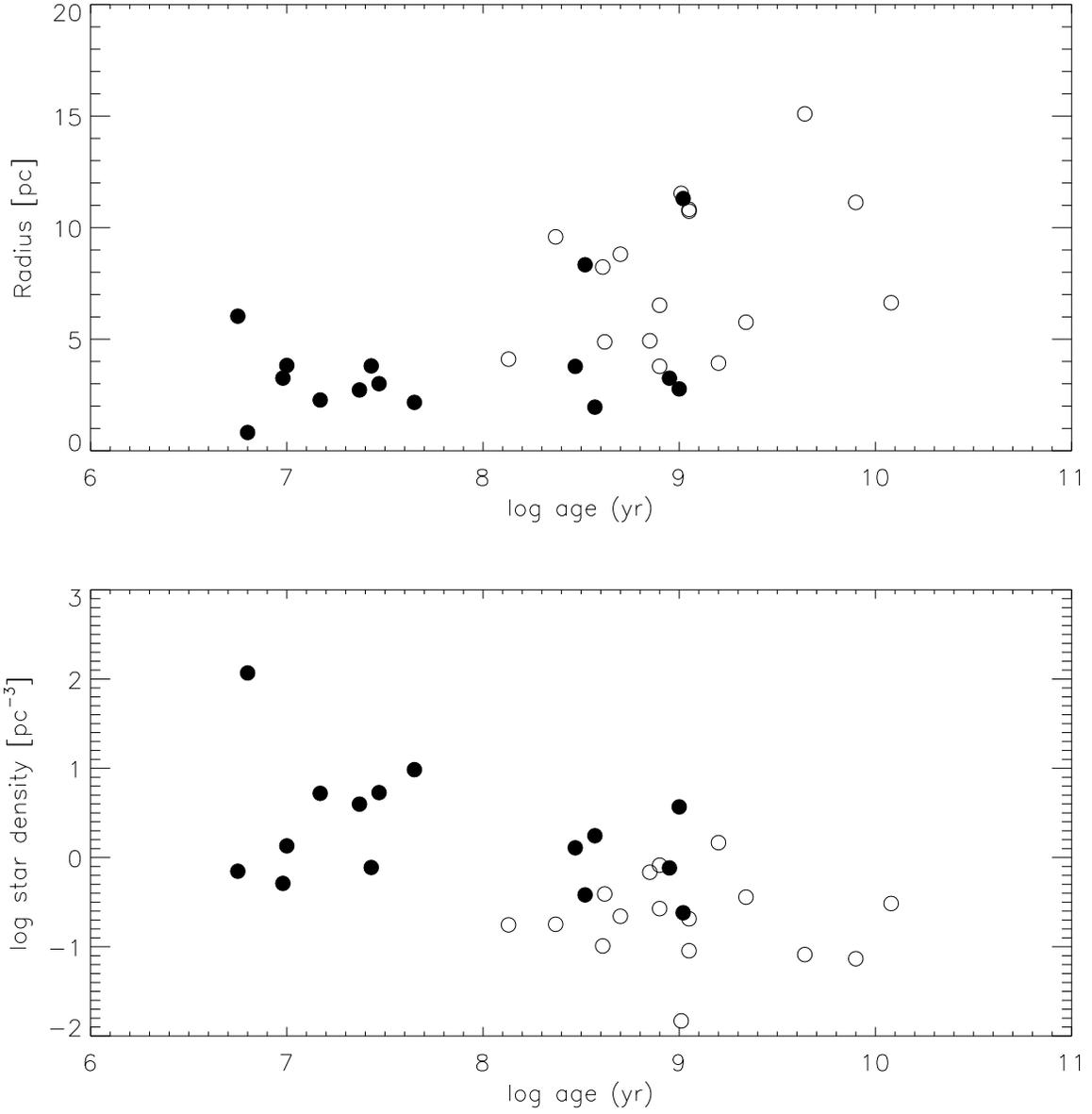} 
 \caption{The physical size (radius in parsec) and richness 
         ($N_{*}$ within the volume, in pc$^{-3}$) of open clusters versus 
          the age.  Older clusters tend to be larger in size and less dense 
          in stellar density.  Filled symbols are for clusters closer to the 
          Galactic plane (cf. Fig.~\ref{fig:ft})   
           }                  
 \label{fig:rdt}
\end{figure}

\begin{figure}
 \plotone{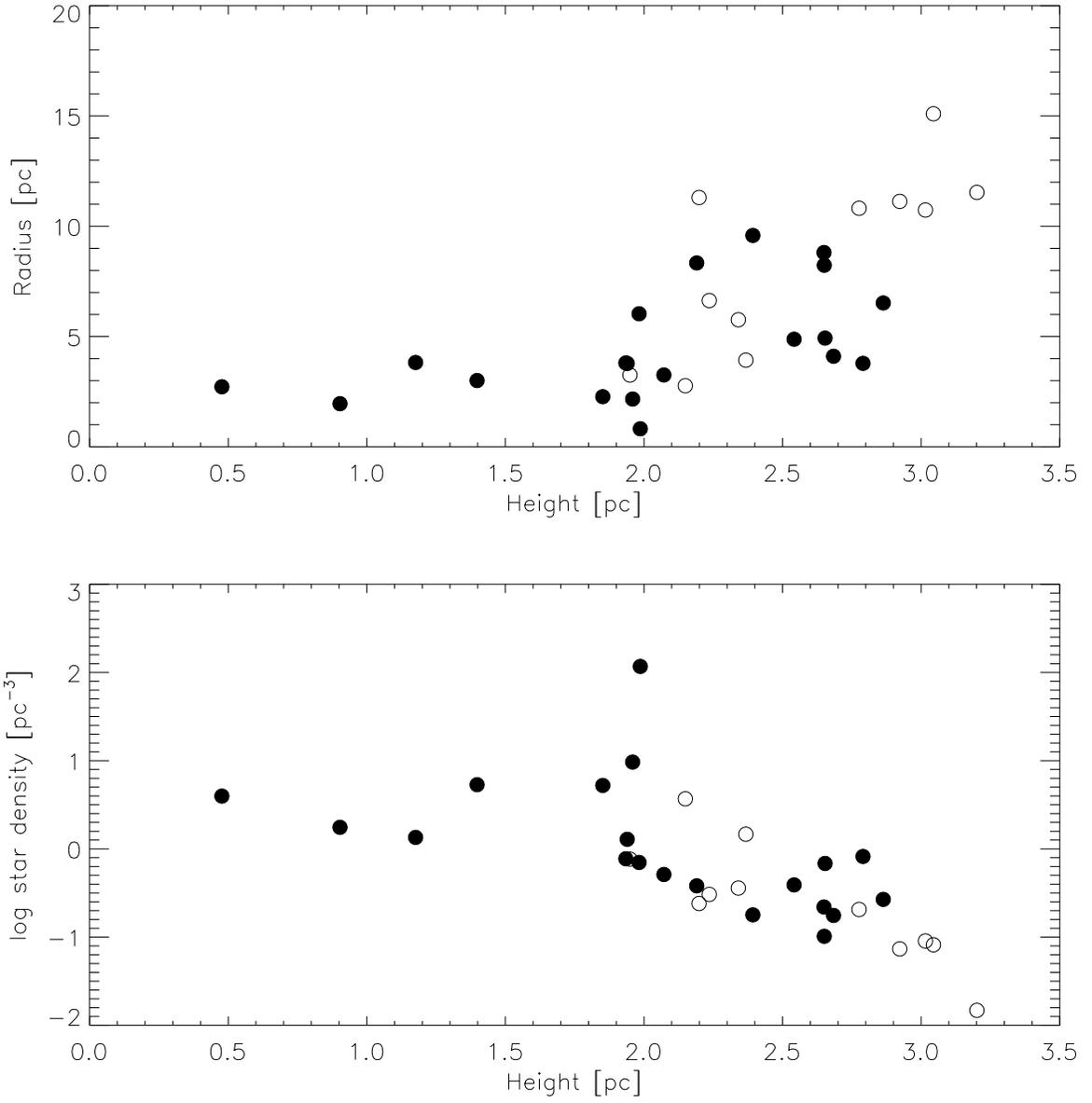} 
 \caption{The physical size (radius in parsec) and richness
         ($N_{*}$ within the volume, in pc$^{-3}$) of open clusters versus
          the height from Galactic disk.  
           }
 \label{fig:rdh}
\end{figure}

Internal stellar dynamics affect not only the shape of a cluster, but 
also its size.  Steller evaporation results in a reduction of number 
of member stars, and the cluster responds by expansion in radius and hence 
a decrease in stellar density.  This is clearly shown in Fig.~\ref{fig:rdt} 
for which older clusters tend to be larger in size and less dense in 
stellar density.  Young open clusters are born to be rich in members, 
amounting to more than $10^{2}$ stars per cubic parsec.  
Subsequent dynamical evolution apparently 
causes them to "loose up" and become dispersed, with the stellar number 
density dropping to that comparable to the field in the solar 
neighborhood, or even lower (for large-height clusters).  
Because old clusters on average have large scale heights, 
it is difficult to distinguish from our data unambiguously the distortion 
effect by aging from that by Galactic tides.  Fig.~\ref{fig:rdh} 
plots the same radius and stellar density of our sample versus 
the height.   The stellar density remains more or less 
the same for clusters within the disk (height less than $\sim 100$~pc), 
yet diminishes markedly with height.  The age seems to play a more 
definite role in the cluster dynamics than Galactic tidal force.

This paper presents our first attempt to delineate the structural evolution 
as a tool to probe the mass distribution and perhaps the dynamical 
evolution of our Galaxy.  Our sample is too small for firm quantitative  
inference, but some preliminary conclusions can be drawn.   
Our study indicates that the shape or morphological 
structure of a young open cluster is dictated by the initial conditions in 
the parental molecular cloud and, as the cluster evolves, by both the internal 
gravitational interaction and external tidal perturbations.  Only the initially 
massive and compact star clusters would have strong enough self-gravitational 
binding to endure the continuing destructive effects, which intensify near the 
disk plane and toward the inner parts of the Galaxy.  Statistics based 
on a larger sample of open clusters are obviously needed to 
quantify, e.g., the time scales, the intricate interplay 
between cluster evolution (age) and its Galactic location 
(galacto-centric distance, and height from the Galactic disk).    

\acknowledgements 
This work makes use of data products from the Two-Micron All-Sky 
Survey (2MASS), which is a joint project of the University of Massachusetts 
and the Infrared Processing and Analysis Center/California Institute 
of Technology, funded by the National Aeronautics and Space 
Administration and the National Science Foundation.  
We thank the referee for very constructive suggestions that greatly 
improve the quality of the paper.  CGS expresses
his gratitude for the hospitality during his visit at NCU.  WPC and CWC
acknowledges financial support of the grant NSC92-2112-M-008-048 
from National Science Council of Taiwan.

\newpage

\begin{deluxetable}{lcrrrcccrcr}
\tablewidth{0pt}
\renewcommand{\arraystretch}{.6}
\tabletypesize\footnotesize
\tablenum{1}
\pagestyle{empty}
\tablecaption{Morphological Parameters of Open Clusters }
\tablehead{
 \colhead{Name} & \colhead{$(l,b)$} & \colhead{$d$} & \colhead{ $H$} & 
 \colhead{$\log$ age}  & 
 \colhead{$ f_{\rm out}$} & \colhead{$f_{1/3}$} &  
 \colhead{$r_{\rm out}$}  & \colhead{$r_{1/3}$} & \colhead{$N_*$}
  \\ 
 \colhead{}  & \colhead{(\arcdeg, \arcdeg)} & \colhead{ (pc)} & \colhead{ (pc) } & 
 \colhead{ (yr)}  &
 \colhead{}               & \colhead{}          & 
 \colhead{(arcmin)}       & \colhead{(arcmin)}  &  
          } 
\startdata
NGC\,2420&(198.11,+19.63) &3085 &1036   &9.05 &0.12 &0.06 &11.60 &3.72 & 468 \\ 
NGC\,2506&(230.56,+09.93) &3460 &597    &9.05 &0.26 &0.15 &10.42 &3.48 & 1091 \\ 
NGC\,1893&(173.59,-01.68) &3280 &-96    &6.75 &0.32 &0.51 &6.13 &4.11  & 645\\ 
King 5&(143.74,-04.27) &1900 &-141      &9.00 &0.43 &0.21 &4.86 &2.65  & 329 \\ 
NGC\,6791&(69.96,+10.90) &5853 &1107    &9.64 &0.20 &0.09 &8.60 &3.90  & 1180 \\ 
Berkeley 17&(175.65,-03.65) &2700 &-172 &10.08&0.51 &0.21 &8.19 &4.23  & 373 \\ 
Melotte 71&(228.95,+04.50) &3154 &247   &8.37 &0.51 &0.12 &10.13 &3.83 & 659 \\ 
NGC\,1245&(146.65,-08.93) &2876 &-446   &8.70 &0.24 &0.02 &10.21 &4.40 & 629 \\ 
Berkeley 69&(174.44,-01.79) &2860 &-89  &8.95 &0.30 &0.28 &3.79 &3.08  & 110 \\ 
NGC\,1960&(174.53,+01.07) &1318 &25     &7.47 &0.32 &0.23 &7.60 &5.65  & 607 \\ 
King 8&(176.39,+03.12) &6403 &348       &8.62 &0.22 &0.28 &2.54 &1.69 & 190 \\ 
Berkeley 21&(186.84,-02.51) &5000 &-219 &9.34 &0.06 &0.14 &3.84 &2.55 & 288 \\ 
NGC\,2414&(231.41,+01.95) &3455 &118    &6.98 &0.11 &0.11 &3.14 &2.81 & 74 \\ 
Trumpler 7&(238.21,-03.33) &1474 &-86   &7.43 &0.40 &0.61 &8.60 &4.56 & 178 \\ 
NGC\,1907&(172.62,+00.31) &1556 &8      &8.57 &0.12 &0.24 &4.19 &3.33 & 55 \\ 
NGC\,2421&(236.27,+00.07) &2181 &3      &7.37 &0.14 &0.12 &4.16 &3.70 & 335 \\ 
NGC\,1817&(186.20,-13.10)  &1972 &-447  &8.61 &0.43&0.37  &13.92 &9.92 & 239 \\ 
NGC\,2567&(249.80,+02.96) &1677 &87     &8.47 &0.44 &0.18 &7.51 &4.02 & 290 \\ 
Ruprecht 18&(239.93,-04.94) &1056 &-91  &7.65 &0.27 &0.22 &6.84 &6.29 & 411 \\ 
NGC\,2354&(238.37,-06.79) &4085 &-483   &8.13 &0.28 &0.23 &3.35 &1.95 & 51 \\ 
Berkeley 39&(223.46,+10.09) &4780 &837  &9.90 &0.18 &0.04 &7.76 &3.55 & 424 \\ 
NGC\,2425&(231.50,+03.30) &4053 &233    &9.20 &0.37 &0.20 &3.23 &2.59 & 372 \\ 
NGC\,2383&(235.27,-02.46) &1655 &-71    &7.17 &0.47 &0.28 &4.58 &2.95 & 258 \\ 
NGC\,2355&(203.39,+11.80) &2200 &450    &8.85 &0.13 &0.21 &7.47 &4.04 & 343 \\ 
NGC\,2158&(186.63,+01.78) &5071 &158    &9.02 &0.11 &0.07 &7.43 &2.89 & 1455 \\ 
NGC\,2194&(197.25,-02.35) &3781 &-155   &8.52 &0.10 &0.21 &7.35 &4.14 & 925 \\ 
NGC\,2204&(226.01,-16.11) &2629 &-730   &8.90 &0.36 &0.26 &8.27 &4.16 & 311 \\ 
NGC\,2304&(197.20,+08.90) &3991 &617    &8.90 &0.17 &0.12 &3.16 &2.20 & 186 \\ 
Tombaugh 2&(232.83,-06.88)&13260 &-1588 &9.01 &0.11 &0.13 &2.90 &2.46 & 95 \\ 
IC\,348&(160.40,-17.72) & 320 &-97      &6.80 &0.47 &0.10 &8.53 &3.87 & 269 \\ 
NGC\,1931&(173.90,+00.28) &3086 &15     &7.00 &0.15 &0.44 &4.13 &2.71 & 316 \\ 
\enddata
\label{tab:results}
\end{deluxetable}


\clearpage

\clearpage

\begin{thebibliography}{} 
%

\bibitem[Adler \& Janes(1982)]{adl82}
  Adler, D. S., \& Janes, K. A., 1982, \pasp, 94, 905

\bibitem[Anthony-Twarog et al.(1990)]{ant90}
  Anthony-Twarog, B. J., Twarog, B. A., Kaluzny, J., Shara, M. M. 1990, 
  \aj, 99, 1504

\bibitem[Bergond, Leon, \& Guibert(2001)]{ber01}
  Bergond, G., Leon, S., \& Guibert, J., 2001, \aap, 377, 462

\bibitem[Binney \& Tremaine(1987)]{bin87}
  Binney, J., \& Tremaine, S., 1987, {\it Galactic Dynamics}, pp. 187 
    (Princeton U. Press)

\bibitem[Chaboyer, Green \& Liebert(1999)]{cha99}
  Chaboyer, B., Green, E. M., \& Liebert, J. 1999, \aj, 117, 1360

\bibitem[Chen, et al.(2003)Chen, Hou, \& Wang]{che03}
  Chen, L., Hou, J. L., \& Wang, J. J., 2003 \aj, 125, 1397

\bibitem[Chen \& Chen(2002)]{che02} 
  Chen, W.P. \& Chen, C.W.  2002, 
           IAU/APRM proceedings

\bibitem[Curry(2002)]{cur02}
  Curry, C. L., 2002, \apj, 576, 849 
 
\bibitem[Danilov \& Seleznev(1990)]{dan90}
  Danilov, V. M., \& Seleznev, A. F., 1990, Bull. Inform. CDS, 
        38, 109

\bibitem[Danilov(1996)]{dan96}
  Danilov, V. M., 1996, IAUS, 174, 389

\bibitem[Dias et al.(2002)]{dia02} 
  Dias, W. S., Alessi, B. S., Moitinho, A., \& Lepine, J. R. D. 2002, 
    \aap, 389, 871, updated information in 
    http://www.astro.iag.usp.br/~wilton/  

\bibitem[Durgapal \& Pandey(2001)]{dur01}
  Durgapal, A. K., \& Pandey, A. K., 2001, \aap, 375, 840

\bibitem[Friel(1995)]{fri95}
  Friel, E. D., 1995, \araa, 33, 381 

\bibitem[Friel et al.(2002)]{fri02}
  Friel, E. D., Janes, K. A., Tavarez, M., Scott, J., Katsanis, R., 
  Lotz, J., Hong, L., Miller, N., 2002, \aj, 124, 2693


\bibitem[Janes \& Phelps(1994)]{jan94}
  Janes, K. A. \& Phelps, R. L., 1994, \aj, 108, 1773 

\bibitem[Jeans(1916)]{jea16}
  Jeans, J. H., 1916, \mnras, 76, 567

\bibitem[Jefferys(1976)]{jef76} 
  Jefferys, W. H., 1976, \aj, 81, 983

\bibitem[King(1961)]{kin61}
  King, I., 1961 \aj, 66, 68

\bibitem[King(1962)]{kin62}
  King, I., 1962, \aj, 67, 471

\bibitem[King(1966)]{kin66}
  King, I., 1966, \aj, 71, 64

\bibitem[Kontizas et al.(1990)]{kon90}
  Kontizas, E., Kontizas, M., Sedmak, G., Smareglia, R., Dapergolas, A., 
  1990, \aj, 100, 425

\bibitem[Kontizas et al.(1991)]{kon91} 
  Kontizas, E, Kontizas, M, Sedmak, G, \& Smareglia, R,   
        1991, IAUS 148, 234

\bibitem[Ka{\l}u\.{z}ny(1994)]{kul94}
  Ka{\l}u\.{z}ny, J. 1994, Acta Astron., 44, 247

\bibitem[Lada \& Lada(2003)]{lad03} 
  Lada, C. \& Lada, E. 2003, \araa, 

\bibitem[Leonard(1988)]{leo88}
  Leonard, P. J. T. 1998, \aj, 95, 108

\bibitem[Lyng\"{a}(1987)]{lyn87a}
  Lyng\"{a}, G., 1987, Catalogue of open cluster data, 
   computer-based catalogue, 5th edition.

\bibitem[Lyng\"{a} \& Palous(1987)]{lyn87b}
  Lyng\"{a}, G., \& Palous, J., 1987, \aap, 188, 35
%


\bibitem[McClure, Forrester, \& Gibson(1974)]{mcc74}
  McClure, R. D., Forrester, W. T., Gibson, J. 1974, \apj, 189, 409

\bibitem[Meyan \& Mayor(1986)]{mey86}
  Meyan, G., \& Mayor, M., 1986, \aap, 166, 122

\bibitem[Nilakshi et al.(2002)]{nil02}
  Nilakshi, N. , Sagar, R., Pandey, A. K., \& Mohan, V., 2002, \aap, 383, 153

\bibitem[Oort (1979)]{oor79}
  Oort, J. H., 1979, \aap, 78, 312

\bibitem[Pandey, Mahra, \& Sagar(1990)]{pan90}
    Pandey, A. K., Mahra, H. S., \& Sagar, R., 1990, \aj, 99, 617 

\bibitem[Phelps et al.(1994)]{phe94}
  Phelps, R. L., Janes, K. A., \& Montgomery, K. A., 1994,
    \aj, 1994, 107, 1079

\bibitem[Portegies Zwart et al.(2001)]{por01}
  Portegies Zwart, S. F., McMillan, S. L. W., Hut, P., Makino, J., 
  2001, \mnras, 321, 199

\bibitem[Salarism, Weiss, \& Percival(2004)]{sal04}
  Salarism, M., Weiss, A., Percival, S. M. 2004, \aap, 414, 163

\bibitem[Tadross(2003)]{tad03}
  Tadross, A. L., 2003, New Astro., 8, 737

\bibitem[Tadross et al.(2002)]{tad02}
  Tadross, A. L., Werner, P., Osman, A., \& Marie, M., 2002, New Astro., 
  7, 553

\bibitem[Vogt \& Moffat(1972)]{vog72}
  Vogt, N., Moffat, A. F. J. 1972, A\&AS, 7, 133

\bibitem[White \& Shawl(1987)]{whi87}
  White, R. E., \& Shawl, S. J., 1987, \apj, 317, 246

\bibitem[Yadav \& Sagar(2001)]{yad01}
  Yadav, R. K. S., Sagar, R. 2001, \mnras, 328, 370 
%
\end{thebibliography}
\end{document}